\documentstyle[11pt]{article}

\newcommand{\be}{\begin{equation}}
\newcommand{\ee}{\end{equation}}
\newcommand{\ba}{\begin{array}}
\newcommand{\ea}{\end{array}}
\newcommand{\bc}{\begin{center}}
\newcommand{\ec}{\end{center}}
\newcommand{\bi}{\begin{itemize}}
\newcommand{\ei}{\end{itemize}}

\newcommand{\disregard}[1]{{}}

\def\bild#1\over#2{\mathrel{\mathop{\kern0pt #1}\limits_{#2}}}

\begin{document}

{\centerline {\bf  EQUATION OF STATE OF AN ANYON GAS
                     IN A STRONG MAGNETIC FIELD \rm}}
\vskip 1cm
{\centerline {\bf Alain DASNI\`ERES de VEIGY and St\a'ephane OUVRY \rm
\footnote{\it  and
LPTPE, Tour 16, Universit\'e Paris  6 / electronic e-mail: OUVRY@FRCPN11}}}

{\centerline {Division de Physique Th\'eorique \footnote{\it Unit\a'e de
Recherche  des
Universit\a'es Paris 11 et Paris 6 associ\a'ee au CNRS},  IPN,
  Orsay Fr-91406}}

\vskip 1cm

Abstract:

The statistical mechanics of an anyon gas in a magnetic field
is addressed. An harmonic regulator is used to define a proper
thermodynamic limit.
When the magnetic field is sufficiently strong,
only exact $N$-anyon groundstates, where anyons occupy the lowest Landau level,
contribute to the equation of state. Particular attention is paid
to the interval of definition of the statistical parameter $\alpha\in[-1,0]$
where a gap exists.
Interestingly enough, one finds that at the critical filling
$\nu=-{1/\alpha}$ where the pressure diverges,
the external magnetic field is entirely screened by the flux tubes
carried by the anyons.

\vskip 3cm

PACS numbers: 05.30.-d, 11.10.-z

IPNO/TH 93-16 (APRIL 1993)

\vfill\eject

-Introduction :
It is now widely accepted that anyons [1] should play a role
in the Quantum Hall effect [2]. In the case of the Fractional Quantum
Hall effect,
Laughlin wavefunctions for the ground state
of $N$ electrons in  a strong magnetic field with filling $\nu=1/m$ provide
an interesting compromise between Fermi degeneracy and Coulomb correlations.
A physical interpretation is that at the critical fractional filling
electrons carry exactly $m$ quanta of flux $\phi_o$ ($m$ odd),
$m-1$ quanta screening the external applied field.
One is left with usual fermions (i.e. anyons carrying one quantum of flux)
in an effective magnetic field  with filling $1$,
or free bosons with no magnetic field.
Anyons with intermediate statistics $1/m$ enter the game
when localized excitations above the groundstate are recognized as carrying
fractional charge and statistics.
The presence of a gap in the spectrum
is crucial for explaining the absence of dissipation on the Hall plateaux.

In the case of the Integer Quantum Hall effect, on the other hand,
one considers a gas of non interacting electrons filling exactely $n$ Landau
levels.
The groundstate is not degenerate,
one has automatically a cyclotron gap
and the Coulomb interaction can be neglected.

In this letter we calculate the equation of state of an anyon gas
in a strong magnetic field.
We argue that considering boson based anyons the $N$-anyon
groundstate problem is entirely solvable in terms of known linear states [3,4],
which end up being product of one-body Landau groundstate.
Particular care is given to
the interval of definition of the statistical parameter $\alpha\in[-1,0]$
in order the gap above the groundstate being under control.
This allows for the analytical derivation of the equation of state
of an anyon gas in a strong magnetic field at low temperature.
We find that the pressure diverges
when the filling factor $\nu$ takes its maximal value
$\nu=-{1/\alpha}$,
suggesting that everything happens as if at most ${-1/\alpha}$ anyons
can occupy a given one-body Landau groundstate\footnote{
after completion of this work,
we noticed that a similar conclusion has been reached in [5],
by a qualitative scaling argument using one-component plasma analogy.
}.
At the critical value of the filling factor,
the anyon gas completely screens the external applied magnetic field,
leaving a free Bose gas.
Moreover, the system is incompressible (both nondegenerate and with a gap).
When $\alpha=-1/n$ where $n$ is an integer,
at the critical filling $\nu=n$
the one-body Landau groundstate has been filled exactly $n$ times,
suggesting a possible reinterpretation of the integer quantum Hall effect
in terms of critical anyons of statistics $-1/n$.

- The model :
Let us  consider in the symmetric gauge the Hamiltonian of
$N$ anyons (charge $e$, flux $\phi$) in a constant magnetic field $B$
($\vec k$ is the unit vector perpendicular to the plane, $\vec r_{ij}=
\vec r_i-\vec r_j$)

\be \label{1}  H_N=\sum_{i=1}^N  {1\over 2m}
               \bigg(\vec{p}_i -\alpha\sum_{j\ne i}
               {\vec k\times\vec r_{ij}\over r_{ij}^2}-e{B\over 2}\vec k
               \times\vec r_i\bigg)^2   \ee
The statistical parameter, $\alpha=e\phi/2\pi$,
measures  the algebraic fraction of quantum of flux $\phi_o=2\pi/|e|$
carried by each anyon. One deals with boson based anyons,
meaning that the wavefunctions $\psi$ are symmetric.
The anyons are coupled to the external magnetic field by their electric
charge $e$.
Coulomb interactions between anyons are ignored. This will be justified a
posteriori when
the anyon gas will be taken at its critical filling
where the groundstate is non degenerate and has a gap.
The Hamiltonian being invariant under
$(x_i,y_i,\alpha,\epsilon)\to (x_i,-y_i,-\alpha,-\epsilon)$, where
$\epsilon=eB/|eB|$,
the spectrum and thus the partition function are invariant under
$(\alpha,\epsilon)\to(-\alpha,-\epsilon)$. They
depend only on $|\alpha|,\epsilon\alpha$ and $\omega_c=
|eB|/2m$ (half the cyclotron frequency). One  chooses $\epsilon=+1$;
in the opposite case
one would  simply change $\alpha\to-\alpha$.
The shift $\alpha\to\alpha+2$ is equivalent to the regular
gauge transformation $\psi\to\exp(-2i\sum_{i<j}\theta_{ij})\psi$,
which does not affect the symmetry of the wavefunctions.
The spectrum is thus periodic in $\alpha$ with period 2.

What is the $N$-anyon groundstate in a magnetic field ?
Let us reexamine this question more closely by paying particular attention to
the domain of definition of $\alpha$. It will turn out that
$\alpha$ has to be taken in the interval $[-1,0]$,
meaning that the magnetic field is antiparallel to the flux tubes
carried by each anyon\footnote{Because of the periodicity $\alpha\to\alpha+2$,
opposite direction has to
be understood to a given even number  quanta of flux.}.

The creation and annihilation operators (in complex coordinate $z_i=x_i+iy_i$)

\be\label{2}   a_i^+ = {1\over\sqrt{m\omega_c}}\left(\partial_i
             -{\alpha\over2}\sum_{j\ne i}{1\over z_{ij}}\right)
             -{\sqrt{m\omega_c}\over2}\bar z_i                  \ee

\be\label{3}    a_i = {1\over\sqrt{m\omega_c}}\left(-\bar\partial_i
                     -{\alpha\over2}\sum_{j\ne i}{1\over\bar z_{ij}}\right)
                     -{\sqrt{m\omega_c}\over2} z_i   \ee
are well defined since they act on anyonic eigenstates
which vanish at coinciding points $z_{ij}=0$.
It follows that they obey the usual commutation rules
$[a_i,a_j^+]=\delta_{ij}$.
The Hamiltonian reads
\be\label{4}    H_N=N\omega_c+2\omega_c\sum_{i=1}^N a_i^+ a_i   \ee
Since the operator $a_i^+ a_i$ is positive,
the lowest energy is $N\omega_c$ and  the $N$-anyon
groundstates are annihilated
by the $a_i$'s. One finds the groundstate basis

\be\label{5}  \psi=z^{l} \prod_{i<j} r_{ij}^{-\alpha}
                   \prod_{i<j} z_{ij}^{m_{ij}}
                   \exp(-{m\omega_c\over 2} \sum_i^Nz_i\bar z_i)
                   \quad\quad l\ge 0 \quad m_{ij}\ge\alpha
\ee
where $z=\sum_i z_i/N$ is the center of mass coordinate.
The total angular momentum is $L=l+\sum_{i<j} m_{ij}$.
Clearly, the $m_{1j}$'s can be choosen as independent orbital quantum numbers.
Moreover, since a bosonic representation has been choosen,
the wavefunctions must be symmetrized,
leading to additional constraints on the $m_{ij}$'s.

If $\alpha$ is not an integer,
the eigenstates (\ref{5}) are entirely contained in the class I [4]
of exact eigenstates of the $N$-anyon problem.
If $\alpha$ is an integer $\alpha_o$,
the $m_{ij}>\alpha_o$ states are contained in class I,
the $m_{ij}=\alpha_o$ states are contained  in class II,
and the remaining states do not belong either to class I or to class II.
The latter states should be obtained at $\alpha=\alpha_o$ from
the non linear states
which are not analytically known.
More precisely,
those states in (\ref{5})  which are not
in class I if $\alpha=\alpha_o$
(i.e. one or more of the $m_{ij}$'s equal to $\alpha_o$)
are not obtained from the states in (\ref{5})
when $\alpha\to \alpha_o$ by superior value.
On the contrary,
when $\alpha\to \alpha_o$ by inferior value,
there is a one to one mapping with the $\alpha=\alpha_o$ states.
This failure to map exactly all states when $\alpha\to \alpha_o$
by superior value is pertinent only when $\alpha_o$ is an even integer
(Bose case).
When $\alpha_o$ is odd (Fermi case), one finds that the states which are
not mapped,
either fall in class I, or simply vanish, after proper symmetrization. Thus
they can
simply be ignored.
This reflects the effect of the exclusion principle
since the remaining states vanish as $r_{ij}^{m_{ij}-\alpha_o}$
with $m_{ij}-\alpha_o\ge 1$ when two particles coincide.

It follows that if one wants to control the gap above
the $N$-anyon groundstate (\ref{5}),
one should accordingly constrain the interval of definition of $\alpha$.
First of all, since we consider boson based anyons,
$\alpha_o$ should be even,
say $\alpha_o=0$ (remember that
the system is periodic in $\alpha$ with period $2$).
Second, one has to consider
anyon eigenstates obtained by negative value $\alpha\le0$.
Finally, the range of variation of $\alpha$ has to end at $-1$,
where the statistics is fermionic.
In this interval, the groundstates (\ref{5}) interpolate continuously
between the bosonic and fermionic groundstates when $\alpha$ decreases from
$0$ to $-1$.
Of course, one should also consider the interval $]-2,-1]$.
However, one  knows for sure that when $\alpha\to -2$ by
superior value some unknown non linear states enter the game,
as the gap between these peculiar states and the groundstate decreases when
$\alpha\to-2$
and finally vanishes at $\alpha=-2$.
Clearly, in this region it is not possible to consider a consistent
thermodynamic of the system in the groundstate.
Semi-classical [6] and numerical analysis [7] for the few-anyon problem
confort this analysis.
In particular, if the semiclassical analysis indicates that
the gap above the groundstate
is indeed of order $2\omega_c$ in the interval $\alpha\in [-1,0]$,
it clearly shows that excited states
merge in the groundstate when
$\alpha\to -2$.

To conclude this discussion, in a regime of strong magnetic field and
low temperature,
where the thermal energy $1/\beta$ is assumed to be smaller than the cyclotron
gap,
the thermal probability $\exp(-2\beta\omega_c)$ to have
an excited state is negligible in the interval $\alpha\in[-1,0]$.
{\it The system is projected into the Hilbert space of the groundstate}.
In fact, if one leaves aside the anyonic prefactor
$\prod_{i<j} r_{ij}^{-\alpha} $,
the $N$-anyon
groundstate basis (\ref{5}) can be rewritten as the direct product
$\{ {\displaystyle \otimes_{i=1}^N} \varphi_{n_i,\ell_i}(z_i)\}_{
n_i=0,\ell_i\ge 0}$
of the  one-body Landau
groundstates
of energy $\omega_c$ and angular momentum $\ell_i$
$$        \varphi_{n,\ell}(z)=z^\ell  L_n^\ell (m\omega_c z\bar z)
          \exp(-{1\over 2} m\omega_c z\bar z)            $$
\be\label{7} \epsilon_{n,\ell}=\omega_c(2n+1) \ee
In this sense, the groundstate of $N$ anyons in a magnetic field
($L=\sum \ell_i$)
is constructed in terms of one-body eigenstate in the lowest Landau level.
As
far as exact anyonic eigenstates are concerned,
the $m_{ij}$'s basis has naturally prevailed. However, when
symmetrizing the Fock space to derive the equation of state, the  $\ell_i$'s
basis will be definitively well adapted.

-The equation of state :
Since one knows exactly the $N$-anyon groundstate spectrum, one can
compute
the $N$-anyon partition function $Z_N$ in the regime
of strong magnetic field
and low temperature.  From the $Z_N$'s one in principle deduces the
cluster coefficients
$b_N$. However, this algorithm happens to be quite tedious
when $N$ becomes large. Instead, one can choose to derive
directly the equation of state in a second quantized formalism
as a power series expansion in $\alpha$. Both methods will be used below.

In order to study the statistical mechanics of an anyon gas,
one should regularize
the system at long distance
to define a proper thermodynamic limit [8].
This is obviously still needed
in the presence of the
magnetic field.
One confines the anyons by a harmonic attraction, adding
$\sum_{i=1}^N {1\over 2}m\omega^2 r_i^2$
to the Hamiltonian (\ref{1}).
The thermodynamic limit is obtained when $\omega\to0$. In the presence
of the harmonic regulator, the
groundstate problem
is simply solved by replacing in (\ref{5})
$\omega_c\to \omega_t=\sqrt{\omega^2+\omega_c^2}$.
The effect of the regulator is to partially lift the degeneracy
of the groundstate spectrum
\be\label{8}  N\omega_c\to N\omega_t+
              \{ l+\sum_{i<j}(m_{ij}-\alpha) \} (\omega_t-\omega_c)
\ee
by $(L-\alpha{N(N-1)/ 2})(\omega_t-\omega_c)$,
where $L-\alpha{N(N-1)/ 2}$ is interpreted as a total orbital angular
momentum
of the $N$-anyon groundstate in the singular gauge.
Again, if one leaves aside the anyonic prefactor, the eigenstates can
be rewritten in terms of one-body
harmonic Landau eigenstates
$$        \varphi_{n,\ell}^{\omega}(z)=z^\ell  L_n^\ell (m\omega_t z\bar z)
          \exp(-{1\over 2} m\omega_t z\bar z)            $$
\be\label{10} \epsilon_{n,\ell}=\omega_t(2n+1) +(\omega_t-\omega_c)\ell
                 \ee
As already stressed above, $N$-anyon states have to be
symmetrized in the case of boson based anyons.
However, as far as symmetry is concerned,
$N$-anyon states and $N$-boson states are identical.
A $N$-anyon state is entirely characterized by the number $n_\ell$
of one-body Landau state of angular momentum $\ell=0,1,...\infty$,
with the constraint $\sum_\ell n_\ell=N$,
and its energy is nothing else but the sum of one-body harmonic Landau levels
$\sum_{\ell}n_\ell\epsilon_{0,\ell}$ shifted by the constant
$-N(N-1)\alpha(\omega_t-\omega_c)/2$.
Since one is interested
by the equation of state, symmetrization is done at the level of
partition functions, in a way quite similar to the bosonic
oscillator equation of state.
The grand partition function for a gas of bosonic oscillators
in the lowest Landau level is
\be\label{11}  Z^b=\prod_{\ell=0}^\infty
                  {1\over{1-z\exp(-\beta\epsilon_{0,\ell})}}
\ee
By definition, $Z^b=\sum_{N=0}^\infty z^N Z_N^b$, where $z$ is the fugacity and
$Z_N^b$ is the $N$-boson partition function.
For a linear spectrum
one can use  the identity $(1-ze^{-\beta\omega_t})Z^b \to Z^b$
when $z\to ze^{-\beta(\omega_t-\omega_c)}$. One deduces that
$Z_N^b -e^{-\beta\omega_t} Z_{N-1}^b
=e^{-N\beta(\omega_t-\omega_c)} Z_N^b$ and finally
\be\label{12}  Z_N^b={e^{-N\beta\omega_t}\over{
                     (1-e^{-\beta(\omega_t-\omega_c)})
                     (1-e^{-2\beta(\omega_t-\omega_c)}) \ldots
                     (1-e^{-N\beta(\omega_t-\omega_c)})        }} \ee
Since the anyonic interaction shifts the $N$-body groundstate spectrum by
$-N(N-1) \alpha (\omega_t-\omega_c)/2$, the $N$-anyon partition function reads
\be\label{13}  Z_N=
               e^{\beta{N(N-1)\over 2}\alpha(\omega_t-\omega_c)}
               Z_N^b  \ee
The thermodynamic limit, $\omega\to 0$, is understood as
$1/(\beta\omega)^2 \to{V/\lambda^2}$, where
the cluster coefficients $b_N$ ($b_2=Z_2-{1\over 2}Z_1^2$,
$b_3=Z_3-Z_2 Z_1+{1\over 3}Z_1^3,\ldots$) are multiplied by $N$ accordingly
[8,9,10]. One infers
\be\label{14}  b_N={V\over \lambda^2}2\beta\omega_c
                {{(N\alpha+1)(N\alpha+2)\ldots (N\alpha+N-1)}\over {N!}}
                e^{-N\beta\omega_c}  \ee
where $V/\lambda^2$ is the volume in unit of the thermal wavelength
$\lambda\equiv \sqrt{2\pi\beta/m}$.

One can see more directly how the volume factor, in the thermodynamic
limit, materializes
in the cluster coefficients using a second quantization  language.
One has to perform a perturbative expansion in $\alpha$ of the thermodynamical
potential. In this context,
short distance singularities
of
the anyon interaction ${\alpha^2\over r_{ij}^2}$ should be treated
before the perturbative analysis can proceed. These singularities
manifest themselves
in the non-analyticity in $|\alpha|$ of the $N$-anyon spectrum, and the fact
that $N$-anyon states have to vanish when two anyons approach each other.
A perturbative
analysis in $\alpha$
is possible [9] if
the $N$-anyon wavefunction is rewritten as
\be
\label{66} \psi(\vec r_1,\cdots,\vec r_N)= \prod_{i<j} r_{ij}^{|\alpha|}
                                     \tilde{\psi}(\vec r_1,\cdots,\vec r_N)
\ee
In (\ref{66}), the exclusion of the diagonal of the configuration space,
a non perturbative effect in $|\alpha|$,
has been encoded, by hand, in the $N$-anyon
wavefunction ($\tilde{\psi}$ is assumed to be non singular).
The Hamiltonian $\tilde H_N$ acting on $\tilde{\psi}$
is known to generate the correct perturbative expansion in $\alpha$ [9,10].
One notes that  the redefinition (\ref{66}) applied to the groundstate
(\ref{5}) precisely factors out the anyonic
prefactor $\prod_{i<j}r_{ij}^{-\alpha}$ in $\tilde{\psi}$.
Thus,  the $\tilde{\psi}$ groundstate
basis does not depend
on $\alpha$, and is identical to the  unperturbed basis.
In the presence of the harmonic regulator $\tilde {H}_N^{\omega}$ reads
$$\tilde {H}_N^{\omega}=     \sum _{i=1}^{N}
              \left[ -{2\over m}\partial_i\bar\partial_i
              +{m\over 2} \omega_t^2 z_i\bar z_i
              -\omega_c (z_i\partial_i-\bar z_i\bar\partial_i)\right]
              \quad\quad\quad\quad\quad\quad $$
\be\label{9}  +\sum_{i<j}\left[-{{|\alpha|-\alpha}\over m}
                {{\bar\partial_i-\bar\partial_j}\over {z_i-z_j}}
              -{{|\alpha|+\alpha}\over m}
                {{\partial_i-\partial_j}\over {\bar z_i-\bar z_j}}
              +\alpha\omega_c\right]                        \ee
When acting  on the groundstate basis
it becomes a sum of one-body Hamiltonian $\sum_i
              -{2\over m}\partial_i\bar\partial_i
              +{m\over 2} \omega_t^2 z_i\bar z_i
              -\omega_c (z_i\partial_i-\bar z_i\bar\partial_i) $
with total energy
$\sum_i\epsilon_{0,\ell_i}$
shifted by
$-\sum_{i<j}\alpha(\omega_t-\omega_c)$.

Second quantizing $\tilde {H}_N^{\omega}$ [10],
the 2-anyon vertex is simply the  constant shift
$-\alpha(\omega_t-\omega_c)/2$.
One uses one particle Green's function
in the lowest Landau level
\be\ba{rl}\label{15}  G_{\beta}(\vec r_2,\vec r_1) & \equiv
            {\displaystyle \sum_{\ell=0}^\infty \varphi_{0,\ell}( z_2)
       \exp(-\beta\epsilon_{0,\ell}) \bar\varphi_{0,\ell}( z_1)} \\ & \\ &=
            {\displaystyle {m\omega_t\over \pi e^{\beta\omega_t}}
             \exp\left(-{m\omega_t\over 2e^{\beta\omega_t}}
             [e^{\beta\omega_c} r_{21}^2+(e^{\beta\omega_t}
             -e^{\beta\omega_c})(r_2^2+r_1^2)+2i
             e^{\beta\omega_c} \vec k.(\vec r_2\times\vec r_1)] \right)} \ea\ee
and computes the diagrammatic expansion of the thermodynamical
potential $\Omega\equiv -\ln\sum_NZ_Nz^N$ as a power series in $\alpha$.
At a given order $\alpha^n$, the leading
connected diagrams (which are the diagrams connected with $n+1$ loops)
are indeed behaving as $1/(\beta\omega)^2$ when
$\omega\to 0$. Also, at this order, non vanishing diagrams start
contributing in the cluster coefficient $b_{n+1}$.
Finally, the thermodynamical potential is found to be
\be\label{16}  \Omega\equiv -\sum_{N=1}^\infty b_N z^N
                    = -{V\over \lambda^2}2\beta\omega_c\ln
y(ze^{-\beta\omega_c})  \ee
where $y(z')$ is solution of $y-z' y^{\alpha+1}=1$ with $y(z')\to 1$
when $z'\to 0$ [11].
The thermodynamical potential
for bosons (fermions) is correctly reproduced when $\alpha=0$ since
$y=1/(1-z')$ (respectively  $\alpha=-1$ since  $y=1+z'$).
The filling factor $\nu\equiv \rho/\rho_L$ (where $\rho_L=2\beta\omega_c
/\lambda^2$ is the Landau degeneracy per unit volume) as a function of $z$ is
given by
$y(ze^{-\beta\omega_c})=1+\nu/(1+\alpha\nu)$. It
is monotically increasing with $z$ from $0$ to $-1/\alpha$.
One deduces the equation of state
\be\label{17}  P\beta=\rho_L
                      \ln\bigg(1+{\nu\over{1+\alpha\nu}}\bigg) \ee
When expanding the pressure as a power series in the density $\rho$, one
verifies that
the
expression of the second virial coefficient
$a_2=-{V\over\lambda^2}{1\over4} {1\over\beta\omega_c} (1+2\alpha)  $
is reproduced
[10,12] in the limit where the Boltzman weight
$\exp(-2\beta\omega_c)$ is neglected\footnote{One finds
$a_3={V\over\lambda^2} {1\over12}
{1\over \beta^2\omega_c^2} (1+3\alpha+3\alpha^2) ,
a_4=\cdots$.}.
Moreover, the first order expansion in $\alpha$ of (\ref{17}) coincides with
the perturbative result in a strong magnetic field given in [10].
The magnetization per unit volume is
\be\label{18}  {\cal M}=-\mu_0\rho+2{\mu_0\over \lambda^2}
                        \ln\bigg(1+{\nu\over{1+\alpha\nu}}\bigg)  \ee
where $\mu_0\equiv |e|/2m$ is the Bohr magneton.
Except near the singularity $\nu=-1/\alpha$,
the
ratio of the logarithmic correction with the De Haas-Van Alphen magnetisation
[13] ${\cal M}=-\mu_0\rho$ is of order $(\beta\omega_c)^{-1}$, and thus
negligible.

Both pressure and magnetization diverge at $\nu=-1/\alpha$.
In the case $\alpha=0$, any value of $\nu$ is allowed
due to Bose condensation.
On the other hand, in the case of Fermi statistics $\alpha=-1$,
Pauli exclusion implies that the lowest Landau level
is completly filled when $\nu=1$.
At a particular $\alpha$, the critical value $\nu=-1/\alpha$
can be interpreted as at most $-1/\alpha$ anyons of statistics $\alpha$
can occupy a given lowest Landau level.
Since transitions to excited levels are by construction
forbidden, the pressure necesseraly diverges when the
lowest Landau level is fully occupied such
that any additional particle is excluded.
In this situation the gas is incompressible.
Indeed  the isothermal compressibility
coefficient $\chi_T=-{1\over V}\left({\partial V\over\partial P}\right)_{T,B}$
vanishes at the critical filling  (except when $\alpha= 0$ where $\chi_T\to
\lambda^2/2\omega_c$).
The groundstate is clearly nondegenerate as can be shown by extracting from
$\Omega$ the canonical partition function of the critical
system $Z_{<N>_{cr}}=
\exp(-\beta <N>_{cr}\omega_c)$ where $<N>_{cr}=V\rho_L(-1/\alpha)$.
Last but not least, one can get some information on the
quantum numbers of the critical nondegenerate  groundstate.
In the fermionic case $\alpha=-1$,
the non degenerate groundstate is known to be a Vandermonde
determinant, built from one body Landau eigentstates $\ell_i=0$  implying
a minimal total angular momentum
$<N>_{cr}(<N>_{cr}-1)/2$
in the singular
gauge. By analogy,  in the case $\alpha\in [-1,0]$ one infers
that  the state
(\ref{5}) with the $\ell_i$'s all equal to $0$ is the critical nondegenerate
groundstate with total angular
momentum $-\alpha <N>_{cr}(<N>_{cr}-1)/2$.

-Discussion :
At the critical filling the magnetic field is entirely
screened by the flux tubes carried by the anyons : each anyon carrying
$\alpha\phi_o$ individual flux, one gets at the critical filling
$\nu=-1/\alpha$
that the flux of the magnetic field is precisely $-N\alpha\phi_o$.

As already emphasized in the introduction, a  similar magnetic screening is at
the origine of the mean-field Chern-Simons-Landau-Ginzburg
theory of the fractional Quantum Hall effect [2].
If $m$ in an odd integer, one can gauge transforms the Hall electrons in boson
based
anyons carrying $m$ quanta of flux.
The mean field solution is meaningful when the
external magnetic field is completely screened
by the flux tubes carried by the anyons. It precisely describes the $\nu=1/m$
fractional Hall liquid.
When $m=1$, one has the $\nu=1$ integer quantum Hall effect where
the lowest Landau level
is entirely filled. In the present situation, each anyon carries $m=-\alpha$
quanta of flux, where $\alpha\in[-1,0]$. At the critical filling
$\nu=-1/\alpha$, again
a magnetic screening occurs; it  corresponds to the  maximum filling of
the lowest Landau
level.

It would certainly be interesting to find out if something special happens
in the particular case $\alpha=-1/n$ with $n$ an integer ($n>1$).
At the critical filling $\nu=n$, the
many-body Landau groundstate has been filled exactly $n$ times. So
one has a non degenerate groundstate, with a cyclotron gap, and an integer
filling $n$.
This suggests a possible reinterpretation
of the $n$ integer quantum Hall effect
in terms of a critical anyon gas of statistics $-1/n$ in a strong
magnetic field with Coulomb interactions  ignored. In the usual
picture of electrons filling $n$ Landau levels, the external magnetic field
is not screened. Here, on the contrary,
the
magnetic field is screened by the critical anyon gas,
quite similarly to the Laughlin wavefunctions in the fractional quantum Hall
effect.

\vfill\eject

\bf {References}\rm

1. J. M. Leinaas and J. Myrheim,  Nuovo  Cimento B {\bf 37}, 1 (1977);
   J. M. Leinaas, Nuovo Cimento A {\bf 47}, 1 (1978);
   M. G. G. Laidlaw and  C. M. de Witt, Phys. Rev. D {\bf 3}, 1375 (1971)

2. R. B. Laughlin, Phys. Rev. B {\bf 23}, 5632 (1981);
   Phys. Rev. B {\bf 27}, 3383 (1983);
   Phys. Rev. Lett. {\bf 50}, 1395 (1983);
   R. Prange and S. Girvin, "The Quantum Hall Effect", Springer, New York
(1987);
   E. Fradkin, "Field Theories of Condensed Matter Systems", Addison-Wesley,
Redwood City (1991);
   A. Lerda, "Anyons", Springer- Verlag (1992);
   S. C. Zhang, Int. J. Mod. Phys. B {\bf 6}, 25 (1992);
   S. Rao, preprint TIFR/TH/92-18

3. Y. S. Wu, Phys. Rev. Lett. {\bf 53}, 111 (1984);
   M. D. Johnson and G. S. Canright, Phys. Rev. B {\bf 41}, 6870 (1990);
   J. Grundberg, T.H. Hansson, A. Karlhede and E. Westerberg,
      ``Landau Levels for Anyons'', preprint USITP-91-2;
   A.P. Polychronakos, Phys. Lett. B {\bf 264}, 362 (1991);
   C. Chou, Phys. Lett. A {\bf 155}, 245 (1991), Phys. Rev. D {\bf 44}, 2533
      (1991);
   K. H. Cho and C. Rim, `` Many Anyon Wavefunctions in a
      Constant Magnetic Field'', preprint SNUTP-91-21;
   A. Khare, J. McCabe and S. Ouvry, Phys. Rev. D {\bf 46}, 2714 (1992);
   L. Brink, T. H. Hansson, S. Konstein and M. A. Vasiliev,
      preprint USITP-92-14;
   A. Govari, "Study of Many Anyons in a
      High Magnetic Field Using a Nonspurious Basis Set";
      "Many Anyons in a High Magnetic Field", preprint Technion-Phy.-92;
   A. Cappelli, C. A. Trugenberger and G. R. Zemba, preprint CERN-TH 6810/93;
   A. Cappelli, G. V. Dunne, C. A. Trugenberger and G. R. Zemba,
      preprint CERN-TH-6784/93

4. G. V. Dunne, A. Lerda and C. A. Trugenberger,
      Mod. Phys. Lett. A {\bf 6}, 2891 (1991);
      Int. Jour. Mod. Phys. B {\bf 5}, 1675 (1991);
   see also G.V. Dunne, A. Lerda , S. Sciuto and C.A. Trugenberger,
      Nucl. Phys. B {\bf 370}, 601 (1992);
   A. Dasni\`eres de Veigy and S. Ouvry,
      Phys. Lett. B {\bf307}, 91 (1993)

5. M. Ma and F. C. Zhang, Phys. Rev. Lett. {\bf 66}, 1769 (1991)

6. S. Levit and N. Sivan, Phys. Rev. Lett. {\bf 69}, 363 (1992);
   J. Aa. Ruud and F. Ravndal, Phys. Lett. B {\bf 291}, 137 (1992)

7. M. V. N. Murthy, J. Law, M. Brack and R. K. Bhaduri, Phys. Rev. Lett.
      {\bf 67}, 1817 (1991);
   M. Sporre, J. J. M. Verbaarschot and I. Zahed,
      Phys. Rev. Lett. {\bf 67}, 1813 (1991);
      `` Four Anyons in an Harmonic Well '', preprint SUNY-NTG-91/40;
      `` Anyon Spectra and the Third Virial Coefficient'',
      preprint SUNY-NTG-91/47;
   A. Khare and J. McCabe, Phys. Lett. B {\bf 269}, 330 (1991);
   J. Myrheim and K. Olaussen, Phys. Lett. B {\bf 299}, 267 (1993)

8. A. Comtet, Y. Georgelin and S. Ouvry, J. Phys. A :
      Math. Gen. {\bf 22}, 3917 (1989);
   K. Olaussen, ``On the Harmonic Oscillator Regularization of
      Partition Function'', Trondheim Preprint (1992)

9. J. McCabe and S. Ouvry , Phys. Lett. B {\bf 260}, 113 (1991);
   see also D. Sen, Nucl. Phys. B {\bf 360}, 397 (1991);
   A. Comtet, J. McCabe and S. Ouvry , Phys. Lett. B {\bf  260}, 372 (1991)

10. A. Dasni\a`eres de Veigy and S. Ouvry, Phys. Lett. B {\bf 291}, 130 (1992);
       ibid, Nucl. Phys. B[FS] {\bf 388}, 715 (1992)

11. E. R. Hansen, "A Table of Series and Products", Prentice-Hall,
       USA (1975) p. 210

12. see M. D. Johnson and G. S. Canright in [3]

13. K. Huang, "Statistical Mechanics", Wiley, USA (1987) p. 260

\end{document}